\documentstyle[12pt,epsf,epsfig,amsmath]{article}
\def\eqright #1\cr{\noalign{\hfill$\displaystyle{{}#1}$}}
\def\eqleft #1\cr{\noalign{\noindent$\displaystyle{{}#1}$\hfill}}

\def\oldreffmt#1{\rlap{[#1]} \hbox to 2\parindent{}}

\def\figfmt#1{\rlap{Figure {#1}} \hbox to 1in{}}

\def\sectioneq{\def\theequation{\thesection.\arabic{equation}}{\let
\holdsection=\section\def\section{\setcounter{equation}{0}\holdsection}}}%

\newcounter{holdequation}

\def\begineq #1\endeq{$$ \refstepcounter{equation}\eqalign{#1}\eqno
	(\theequation) $$}
\def\contlimit{\,{\hbox{$\longrightarrow$}\kern-1.8em\lower1ex
\hbox{${\scriptstyle (a\rightarrow0)}$}}\,}
\def\centeron#1#2{{\setbox0=\hbox{#1}\setbox1=\hbox{#2}\ifdim
\wd1>\wd0\kern.5\wd1\kern-.5\wd0\fi
\copy0\kern-.5\wd0\kern-.5\wd1\copy1\ifdim\wd0>\wd1
\kern.5\wd0\kern-.5\wd1\fi}}
\def\centerover#1#2{\centeron{#1}{\setbox0=\hbox{#1}\setbox
1=\hbox{#2}\raise\ht0\hbox{\raise\dp1\hbox{\copy1}}}}
\def\centerunder#1#2{\centeron{#1}{\setbox0=\hbox{#1}\setbox
1=\hbox{#2}\lower\dp0\hbox{\lower\ht1\hbox{\copy1}}}}
\def\lsim{\;\centeron{\raise.35ex\hbox{$<$}}{\lower.65ex\hbox
{$\sim$}}\;}
\def\gsim{\;\centeron{\raise.35ex\hbox{$>$}}{\lower.65ex\hbox
{$\sim$}}\;}

\def\super#1{\ifmmode \hbox{\textsuper{#1}}\else\textsuper{#1}\fi}
\def\textsuper#1{\newcount\holdspacefactor\holdspacefactor=\spacefactor
$^{#1}$\spacefactor=\holdspacefactor}

\def\getcite#1,{\advance\citenumber by1
\def\getcitearg{#1}\def\lastarg{@}
\ifnum\citenumber=1
\ref{#1}\let\next=\getcite\else\ifx\getcitearg\lastarg\let\next=\relax
\else ,\ref{#1}\let\next=\getcite\fi\fi\next}

\def\pom{{\rm P\kern -0.53em\llap I\,}}
\def\spom{{\rm P\kern -0.36em\llap \small I\,}}
\def\sspom{{\rm P\kern -0.33em\llap \footnotesize I\,}}

\relax
\def\contlimit{\,{\hbox{$\longrightarrow$}\kern-1.8em\lower1ex
\hbox{${\scriptstyle (a\rightarrow0)}$}}\,}
\def\upon #1/#2 {{\textstyle{#1\over #2}}}
\relax
\renewcommand{\thefootnote}{\fnsymbol{footnote}}

\sectioneq

\def\til#1{\centeron{\hbox{$#1$}}{\lower 2ex\hbox{$\char'176$}}}
\def\tild#1{\centeron{\hbox{$\,#1$}}{\lower 2.5ex\hbox{$\char'176$}}}
\def\sumtil{\centeron{\hbox{$\displaystyle\sum$}}{\lower
-1.5ex\hbox{$\widetilde{\phantom{xx}}$}}}

\begin{document} 

\begin{titlepage} 

\rightline{\vbox{\halign{&#\hfil\cr
&\today\cr}}} 
\vspace{0.25in} 

\begin{center} 
  
{\large\bf High-Energy Unitarity and the Standard Model}\footnote{Work 
supported by the U.S.
Department of Energy under Contract
W-31-109-ENG-38} 

\medskip

Alan. R. White\footnote{arw@hep.anl.gov }

\vskip 0.6cm

\centerline{Argonne National Laboratory}
\centerline{9700 South Cass, Il 60439, USA.}
\vspace{0.5cm}

\end{center}

\begin{abstract}
 
High-energy unitarity is argued to select a special version
of QCD as the strong interaction. Electroweak symmetry breaking has to be 
due to a new sextet quark sector - that will produce large cross-section effects 
at the LHC. The sextet sector embeds, uniquely, in a massless $SU(5)$ theory that 
potentially generates the states and interactions of the Standard Model 
within a bound-state S-Matrix.
Infra-red chirality transitions of the massless Dirac sea play an essential
dynamical role. 

\end{abstract} 

\vspace{2in}

\centerline{Contributed to the Proceedings of the Gribov-75 Memorial Workshop.}

\renewcommand{\thefootnote}{\arabic{footnote}} \end{titlepage}

\section{Introduction.}

Vlodya Gribov devoted much of his life to studying the implications of
unitarity for a high-energy S-Matrix. My own work is based heavily on Vlodya's 
ideas and formalisms and in this article, written in honor of 
what would have been his 75th birthday, I will outline results suggesting that
high-energy unitarity could be a determining constraint on a particle 
S-Matrix. 
I will first argue that a special version of QCD (QCD$_S$), with experimentally 
attractive features, is selected as the strong interaction\cite{arw04}. 
A new color sextet quark sector is present that has just the right properties  
to produce electroweak symmetry breaking - with large cross-section effects 
predicted for the LHC.  A unique embedding\cite{kw} of the sextet sector 
in a left-handed $SU(5)$ theory (GUT$_S$)
removes the sextet electroweak anomaly and also, 
amazingly, includes ``almost exactly'' 
the lepton and triplet quark sectors of the Standard Model, with 
QCD$_S$ included in it's entirety. Because some of the $SU(2)\otimes U(1)$ 
quantum numbers are not quite right, GUT$_S$ can 
not be a conventional unified theory. Rather, it may  
provide an underlying{ \it{massless}} field theory which dynamically
generates, {\it within the bound-state S-Matrix,}
both the triplet and sextet hadronic sectors and, also,
the leptonic sector of the Standard Model. 
If this proves\footnote{Most importantly, 
sextet electroweak symmetry breaking must be seen\cite{arw04} at the LHC.} 
to be the case then (it will be arguable that) unitarity uniquely 
determines the Standard Model ! 

My results imply that the infra-red chiral anomaly effects of a massless 
Dirac sea can have a deep significance in a bound-state S-Matrix. 
At  high-energy, zero momentum chirality transitions
(within reggeon vertices) produce states and amplitudes with
dramatically different properties  
from those implied, at first sight, by the underlying field theory. In the 
S-Matrix of massless QCD$_S$, {\it the chirality transitions produce 
confinement and chiral symmetry breaking} - with
a minimal hadronic (triplet and sextet) spectrum\cite{arw04},
and Critical Pomeron high-energy behavior\cite{cri}. Although much remains to 
be understood about the S-Matrix of GUT$_S$, it is already clear that, {\it  
just because of the Dirac sea interactions, the Standard Model S-Matrix
could indeed emerge}. The $SU(5)$ symmetry is confined 
and so, very importantly, proton decay is not directly implied. In addition, 
because the chirality transitions occur only for vector coupled fermions, 
{\it the real $SU(3)\otimes U(1)$ part of the theory dominates the S-Matrix}. 
A resulting ``wee gluon anomaly interaction'' gives\cite{arw04} a mass 
for SU(3) singlet left-handed vector bosons and 
the full interaction structure of the
Standard Model appears. There is a related bound-state mass spectrum in which, because of
the fermion representation structure, there are no (unwanted) symmetries.

I will emphasize where Gribov's work underlies the 
arguments that I present, 
beginning with the formulation of reggeon unitarity\cite{gpt,arw00} and ending 
with the importance of the dynamical role\cite{yd} of the Dirac sea.
The Critical Pomeron solution\cite{cri} of 
reggeon unitarity and the reggeization of 
non-abelian gauge theories (that has been established, in remarkable depth, by 
Lev Lipatov and collaborators\cite{fkl}) are also crucial.

\section{Reggeon Unitarity }

The multi-regge behavior of scattering amplitudes is determined by partial-wave 
amplitudes analytically continued in complex angular momentum. The singularity 
structure in each ``J-plane'' is controlled by $t$-channel unitarity. Elastic
unitarity generates regge poles which, in turn, generate regge cuts via
multiparticle unitarity. The reggeon unitarity equations then determine the threshold 
discontinuity due to any combination of regge poles, in any partial-wave amplitude. 
If $\alpha(t)$ is a regge pole trajectory, the $M$-reggeon cut trajectory is 
$ J=\alpha_M(t)= M\alpha(t/M^2) - M +1$.
 
When the reggeon unitarity equations were first derived\cite{gpt},
they were a spectacular intellectual jump 
and a profound reformulation and generalization of existing low-order diagrammatic
calculations. There were, however,
many uncertainties in the derivation that were resolved only after
the development\cite{arw00} of multiparticle symptotic
dispersion relations as a fundamental basis for multi-regge theory. The
generality of reggeon unitarity makes it 
extremely powerful, particularly when applied to the problem of constructing 
the multi-regge region QCD S-Matrix. First we 
discuss the abstract Reggeon Field Theory 
(RFT) solution of these equations.

\section{The Critical Pomeron}

The analogy with conventional unitarity makes it straightforward to formulate
an effective field theory solution of reggeon unitarity. In fact,
Gribov developed\cite{gr} a direct (underlying) diagrammatic formulation of RFT only 
because\cite{arw00} of the uncertainties that, for a long time, surrounded the 
derivation of reggeon unitarity. We consider an, even-signature,
pomeron regge pole with trajectory $j = \alpha(t)$. If $\alpha(0) = 1$ then, 
also, $\alpha_M(0) = 1$ and
all the multipomeron singularities accumulate at one point. In this case, 
a simultaneous solution of all the discontinuity formulae is essential and all RFT
diagrams have to be summed. Remarkably, a renormalization group formalism can be 
applied and a fixed-point solution shown to exist.

Under a renormalization group transformation, which rescales both $(1-J)$ 
and $k_{\perp}$,
only the triple pomeron coupling $r$ survives as a relevant coupling. 
A fixed-point exists at $\alpha(0)=1$ and $r^2=\frac{4\pi^2}{3} \epsilon
+O(\epsilon^2)$
(where $\epsilon=4-D$ and $D$ is the dimension of $k_{\perp}$). 
$\epsilon=2$ gives an interacting pomeron theory with
the ``universality'' property of a critical 
phenomenon. As a result, the asymptotic behavior 
can be precisely calculated without knowledge of the underlying ``bare''
parameters. Scaling laws can be derived 
for a wide variety  of cross-sections and it can be shown
that all known $s$- channel unitarity constraints on 
a theory of rising cross-sections are satisfied\cite{mm}. It can also be shown that, 
when $t$ is positive, all the multipomeron cuts separate and reggeon unitarity 
is satisfied. Therefore, the Critical Pomeron provides a complete, unique, 
solution of high-energy unitarity which, in a sense, represents the ultimate 
achievement of the pre-QCD program to construct a unique S-Matrix
from unitarity. {\it (It has been my goal, for a long time, to determinine how 
and when the Critical Pomeron occurs in a field theory.)}

\section{The Supercritical Pomeron}

The supercritical phase\cite{arw91} (obtained when, formally, $\alpha(0)>1$)
plays a crucial role in linking
the Critical Pomeron to an underlying field theory. Using 
a stationary point of the effective lagrangian introduces a ``pomeron condensate'' 
that generates new classes of RFT diagrams. 
Figure 1 contains a simple example of a new diagram. 
\begin{figure}[ht]
\centerline{\epsfxsize=3.5in\epsfbox{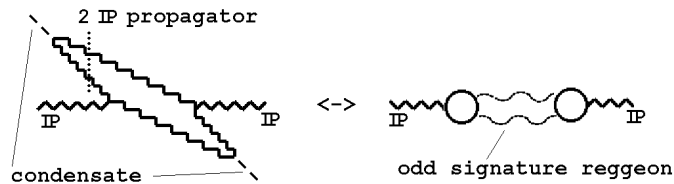}}   
\caption{ A new RFT diagram generated by the pomeron condensate}
\end{figure}
Reggeon unitarity determines that the $k_{\perp}$ poles produced by the
two-pomeron propagators coupled to the condensate
should be interpreted as particle poles lying on an odd-signature 
trajectory degenerate with that of the pomeron. The odd-signature reggeon 
couples pairwise to the pomeron and reggeon states involving many 
vector particle poles similarly appear in higher-order diagrams.
In the supercritical phase, therefore, 
divergences in rapidity (produced by $\alpha(0)> 1$)
are converted to vector particle divergences in 
$k_{\perp}$ that are associated with the ``deconfinement of a vector particle'' on the
pomeron trajectory. 

\section{The Critical Pomeron and QCD$_S$.}

If the Critical Pomeron is not present in a theory then, since perturbative 
cross-sections rise, it is unlikely that large momentum 
perturbation theory can match smoothly with unitary forward amplitudes.
In fact, before even discussing dynamical details, we can give some general arguments 
that specifically link the Critical Pomeron to a special version of QCD (QCD$_S$).
 
Breaking $SU(3)$ color to $SU(2)$ gives ``color superconducting QCD'' (CSQCD) in which
a massive, reggeized, vector particle 
(a ``massive gluon'') is deconfined, just as is expected in the supercritical pomeron
phase\footnote{That the gauge group has to be $SU(3)$ can be seen\cite{arw80}, 
without the detailed construction of the next
Section, by using cut RFT to determine the center of the group.}. 
However, at large momentum, a smooth 
transition from CSQCD to QCD is possible only if 
an asymptotically free scalar ``Higgs'' field can be employed. This is a strong
requirement that is only possible\cite{arw04,gw} 
if the asymptotic freedom constraint on QCD is
``saturated'' - immediately implying that there must be a further quark sector 
beyond that present in the conventional version of the Standard Model.
Also, in the infra-red transverse momentum region, the transition from CSQCD to QCD 
can give the Critical Pomeron only if
the infra-red divergences of perturbative reggeized gluon diagrams 
can generate 
non-perturbative pomeron diagrams describing a confining 
theory. As we will see, 
an infra-red fixed-point, that
again exists only when asymptotic freedom is saturated,
is crucial - in addition to chiral anomalies.

The saturation constraint (unrealistically) requires sixteen color triplet
quarks or, alternatively, two 
color sextets and six triplets (giving QCD$_S$). The resulting 
``sextet pions'' provide, exactly, a ``Higgs sector'' producing electroweak 
mass generation\cite{arw04,wm}. Thus, 
requiring unitarity for the strong interaction leads directly to electroweak 
symmetry breaking and implies that there is a major sector of the 
strong interaction that has yet to be seen experimentally.

\section{CSQCD$_S$ States and Amplitudes}

The derivation of supercritical pomeron diagrams required\cite{arw91} the study of
multi-regge amplitudes in which 
the ``pomeron condensate'' could be understood as due to 
a ``wee parton'' component of scattering hadrons. Correspondingly, to derive 
supercritical states and amplitudes within CSQCD$_S$, we have to consider a multi-regge 
limit\cite{arw98} in which the amplitude for the scattering 
of regge pole hadrons via pomeron exchange can emerge in it's entirety,
as illustrated schematically in Figure 2. 
\begin{figure}[ht]
\centerline{\epsfxsize=0.4in\epsfbox{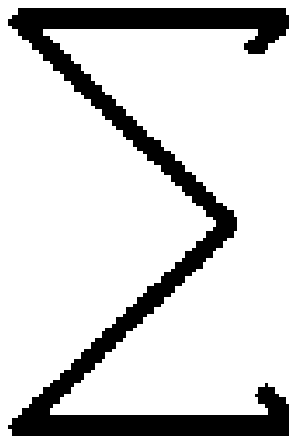}\epsfxsize=2.4in\epsfbox{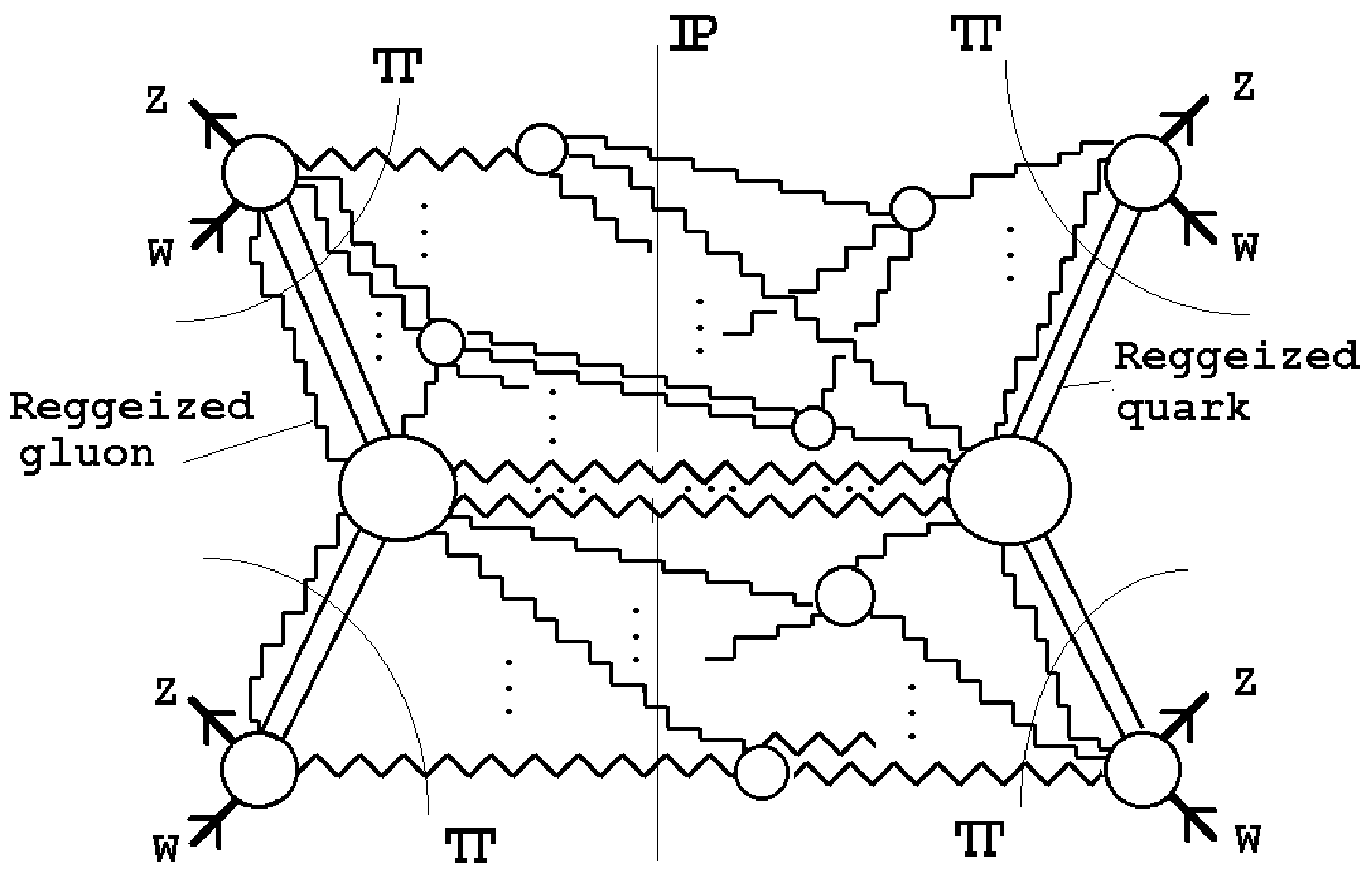}
\epsfxsize=2.3in\epsfbox{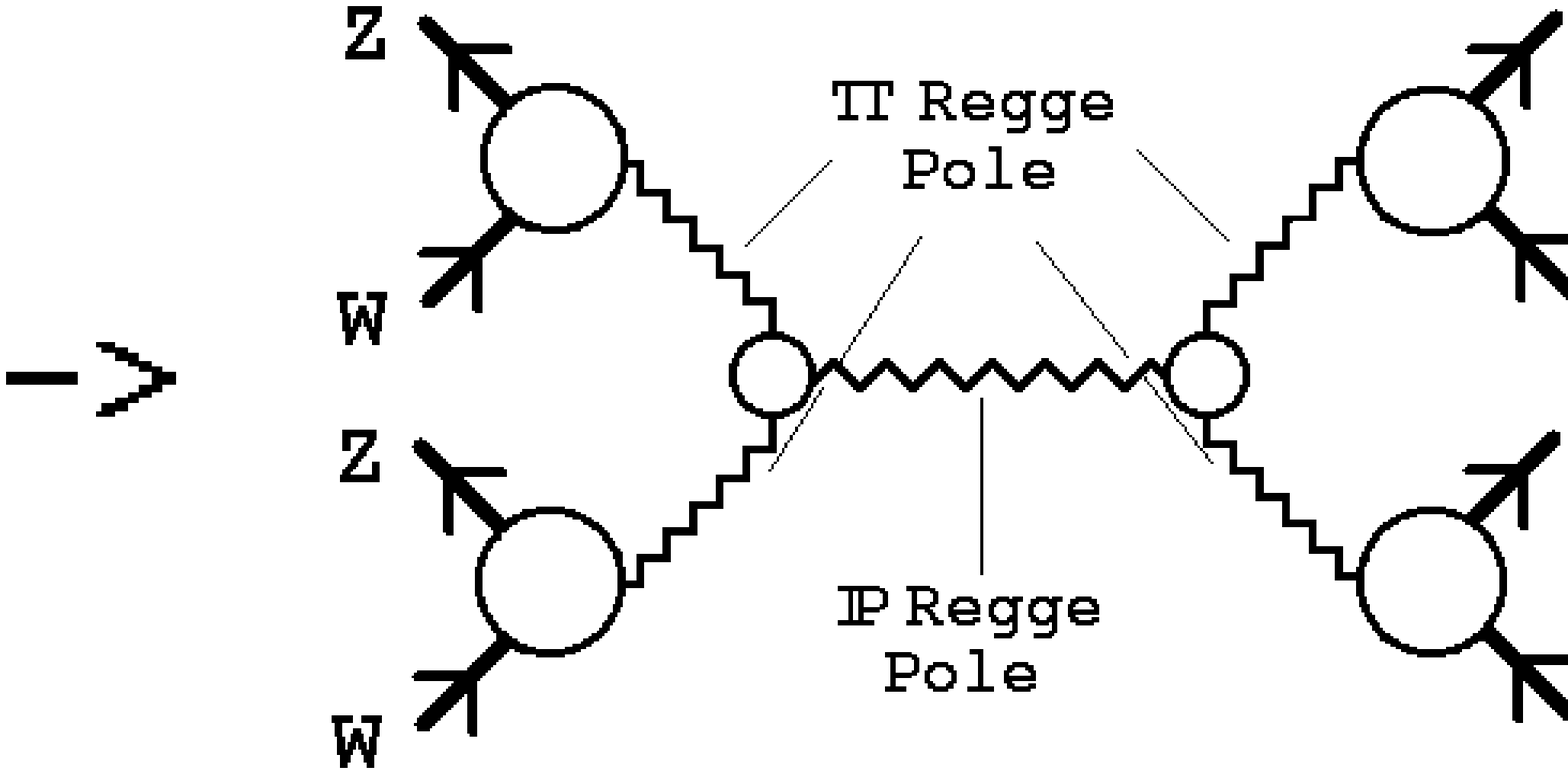}}   
\caption{ The transition from perturbative reggeon diagrams to reggeized 
pions scattering via pomeron exchange.}
\end{figure}
Fortunately, the generality of 
reggeon unitarity implies that the complicated reggeon diagrams 
involved can be constructed similarly to the well-known elastic scattering
diagrams, the only difference being that vertices coupling distinct
reggeon channels can contain anomalies not present in internal
reggeon interactions.

The main infra-red divergence of massless gluon reggeon diagrams 
is due to reggeization. This divergence
exponentiates to zero all amplitudes with non-zero $SU(2)$
color in any reggeon channel.
We consider diagrams, of the form shown in Figure 3, 
\begin{figure}[ht]
\centerline{$~$\hspace{0.1in}\epsfxsize=2.1in\epsffile{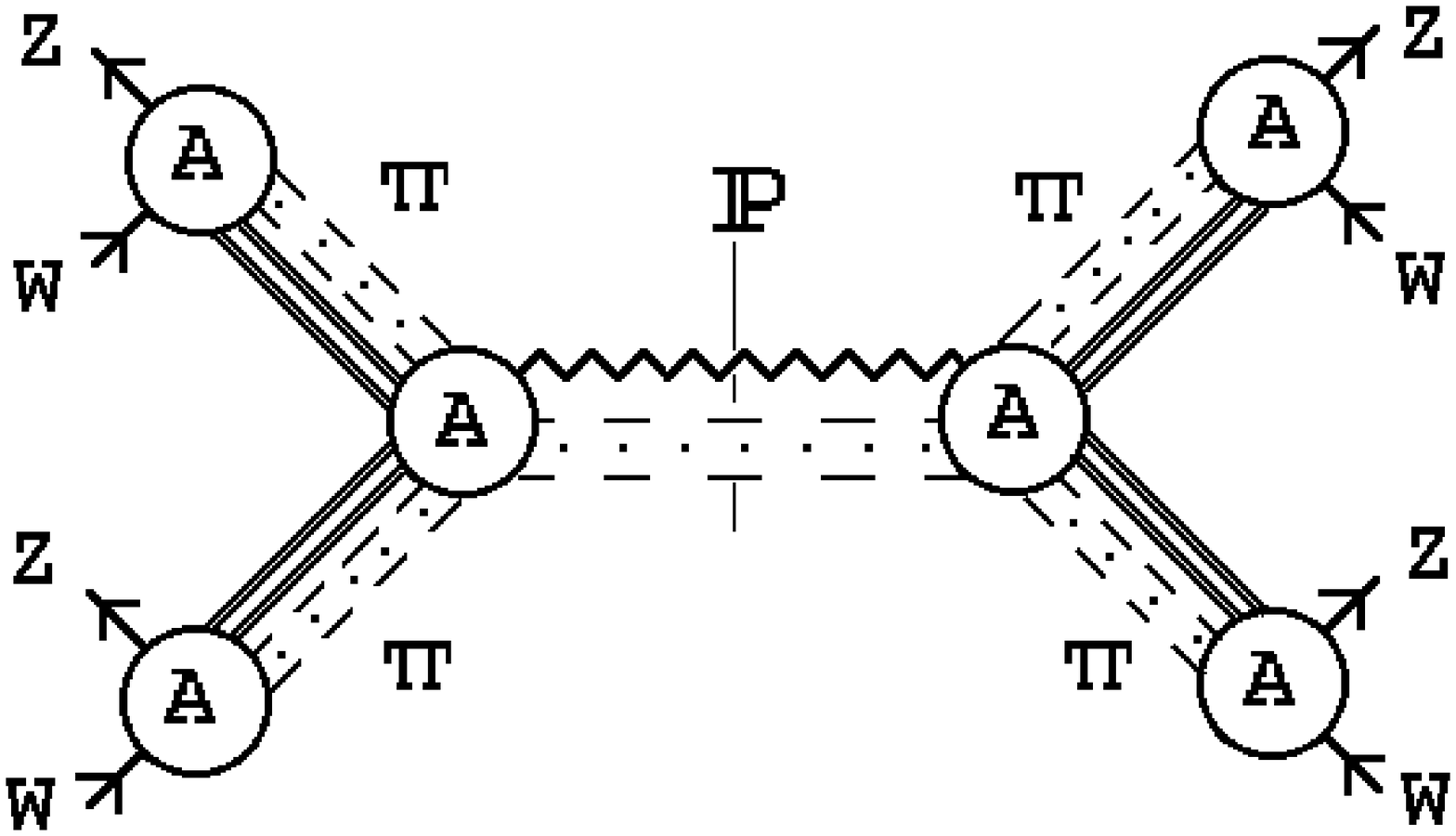}
\hspace{0.3in}\epsfxsize=2.4in
\epsfbox{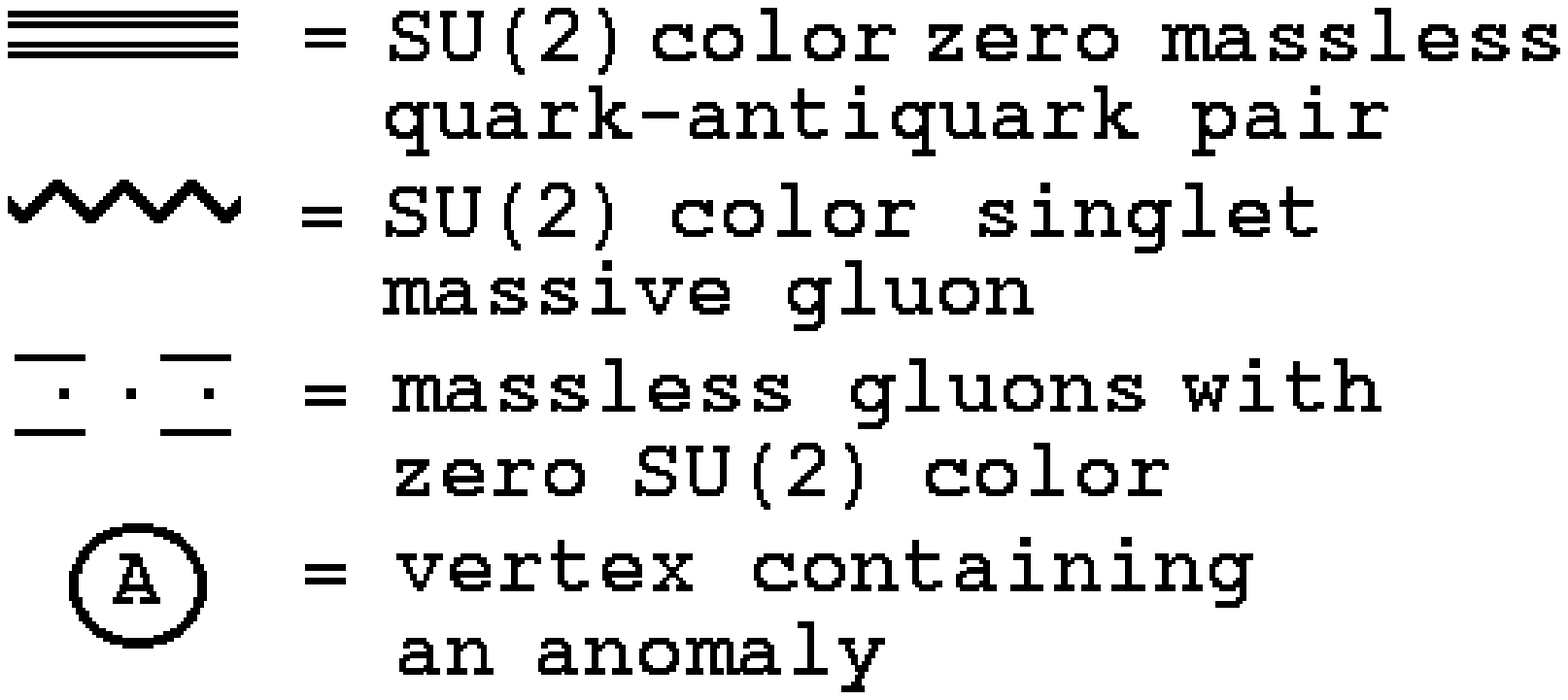}}   
\caption{Diagrams with anomaly vertices that will produce the pion amplitude.}
\end{figure}
in which each
reggeon state has two components, both of which carry 
zero $SU(2)$ color.
One component contains massless gluons 
while the other is either a  
quark-antiquark pair or a massive gluon reggeon.
We choose external left-handed vector boson vertices 
because, as illustrated in Figure 4, they directly 
\begin{figure}[ht]
\centerline{\epsfxsize=4.8in\epsfbox{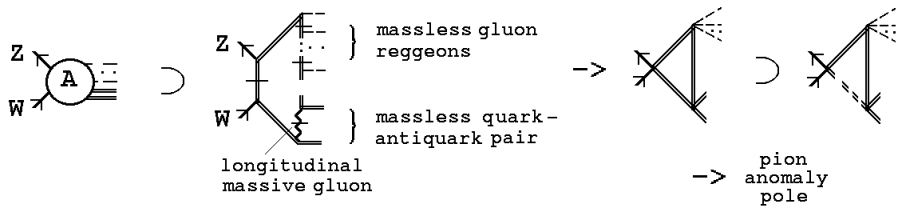}}   
\caption{A reggeon vertex triangle diagram. The hatched
lines are on-shell and the broken quark line
indicates a zero momentum chirality transition.
} 
\end{figure}
contain\cite{arw03} a triangle anomaly.
Consequently, an ``anomaly pole'' appears, as shown, 
when the gluon reggeons carry zero transverse momentum.
The internal vertices in Figure 3
contain U(1) anomaly diagrams of the form shown in Figure 5.
\begin{figure}[ht]
\centerline{\epsfxsize=4in
\epsfbox{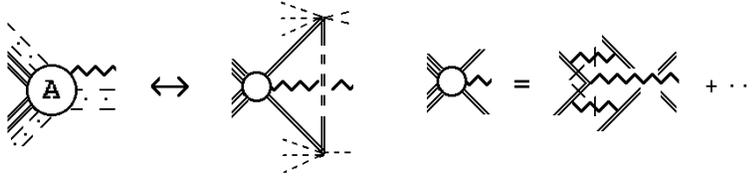}}   
\caption{The reduction to a triangle diagram that involves the U(1) anomaly.
The broken quark line again indicates the chirality transition giving
the anomaly pole.}
\end{figure}

The initial effect of the anomalies 
is a large transverse momentum (non-unitary) power
enhancement\cite{arw03} of the high energy behavior. The enhancement 
can be removed with a transverse momentum cut-off but, 
because gauge invariance Ward identities are violated,
a new infra-red divergence is produced.
An overall (logarithmic) divergence is generated when 
all massless gluon transverse momenta are scaled uniformly 
to zero. (The $U(1)$ anomaly pole contributes as a 
transverse momentum conserving $\delta$-function.) However, when the
massless gluon component in any reggeon state carries normal color 
parity ($= $ signature), interactions with the second component in the state
again exponentiate the divergence.
For anomalous ($\neq $ signature) color parity gluon components, these interactions 
are absent.

The infra-red fixed point, due to saturation, implies that
color zero massless gluon reggeon interactions
scale\footnote{This is  
related to the conformal invariance 
of Green's functions at the fixed-point.}  
as all transverse momenta are scaled to zero. 
The overall ``anomalous wee gluon'' divergence
is then preserved when interactions amongst the massless gluons are included. 
The residue of 
the divergence gives a physical CSQCD$_S$ amplitude, in which the transverse
momentum cut-off can be removed and all the
massless gluons in Figure 3 contribute only as a ``reggeon condensate''. The 
condensate provides (crucial) zero transverse momentum, anomalous wee gluon,
components in both the ``pion'' and the ``pomeron''. 

$SU(2)$ anomalous gluons must have odd signature and so
the CSQCD$_S$ pomeron is an even signature regge pole which, because of the
reggeon condensate, is
exchange degenerate with a reggeized massive gluon, just as in 
supercritical RFT. Also because of the condensate, ``pion'' 
Goldstone boson anomaly poles\footnote{Via the chirality transition, the
condensate is absorbed by a zero momentum anti-quark (or quark) 
that becomes unphysical, as in Gribov's confinement picture\cite{yd}.}, 
from the external vertices, appear in the
physical amplitudes. By removing the transverse 
momentum cut-off {\it{after the extraction of anomaly
infra-red divergences}}, we replace ultra-violet 
chirality violation (producing bad
high-energy behavior) by infra-red chirality violation  
producing particle poles. This is how a confining, chiral symmetry breaking, 
bound-state spectrum is generated out of reggeon diagrams. 
The spectrum consists of triplet and sextet 
quark/antiquark ``pions'' and, due to the equivalence of conjugate
$SU(2)$ representations, tripet and sextet quark/quark (and anti-quark/anti-quark)
``nucleons''.

\section{Restoration of $SU(3)$ color}

There is every indication\cite{arw04} that the reggeon diagrams of $CSQCD_S$ 
map onto supercritical RFT, with the essential triple pomeron coupling 
given\cite{arw02} by wee gluon anomaly interactions. It remains, however,
a major challenge to carry out the mapping in detail.
Assuming it can be done, the transition from CSQCD$_S$
to $QCD_S$ will necessarily give the Critical Pomeron.
The reggeon condensate will disappear and  
the zero momentum chirality transitions (which can, equivalently, be regarded
as Dirac sea shifts) in states and amplitudes 
will no longer be due to a semi-classical gauge field
with fixed $SU(3)$ color. The transitions will remain and
will be many in any scattering process, but they will 
correspond to random, dynamical, gauge field fluctuations within the color group. The 
transition from a fixed ``magnetization'' for the Dirac sea shifting
gauge field, to a random, fluctuating, field characterizes the
nature of the ``critical phenomenon'' that is associated with the high-energy 
behavior of $QCD_S$. In effect, the longitudinal massive
vector meson interactions do not decouple entirely as the 
color symmetry breaking is removed. They remain as dynamical fluctuations, but 
only at zero light-cone momentum. Including such interactions (specifically) 
amounts to fixing the (Gribov) ambiguity in the light-cone quantization of QCD$_S$.

The quarks clearly have to be massless for the 
physics of QCD$_S$ to be as we have described it. To add
quark masses, and preserve the
physics involved, appears non-trivial. 
The Dirac sea would have to undergo major shifts,
of the kind envisaged by Gribov in his original confinement proposal\cite{yd},
but in a random dynamical manner.
In fact, the solution may be that effective quark masses are introduced via
the bound-state masses originating from the embedding of $QCD_S$ in GUT$_S$, 
that we discuss below.

QCD$_S$ baryons will be formed as bound states of CSQCD$_S$ nucleons
and $SU(2)$ singlet quarks. Crucially, because there are no chiral 
symmetries mixing the two sectors,
there will be no ``hybrid'' sextet/anti-triplet/anti-triplet
combinations. Therefore, the only new baryons will be 
the sextet proton and the (stable) sextet neutron. 

\section{ The Sextet QCD Scale and Electroweak Masses }

Sextet antiquarks have the same $SU(3)$ triality as triplet quarks
and so we define ``Standard Model'' couplings for sextet antiquarks (quarks) to be 
the same as triplet quarks (antiquarks). In an infinite momentum hadron, 
wee gluons should reproduce vacuum properties
and, indeed, an anomaly interaction
generates a mass\footnote{Similar interactions
should also generate a pion mass in the even signature amplitude.} 
for an exchanged vector boson,
as illustrated in Figure 6. 
\begin{figure}[ht]
\centerline{\epsfxsize=1.2in
\epsfbox{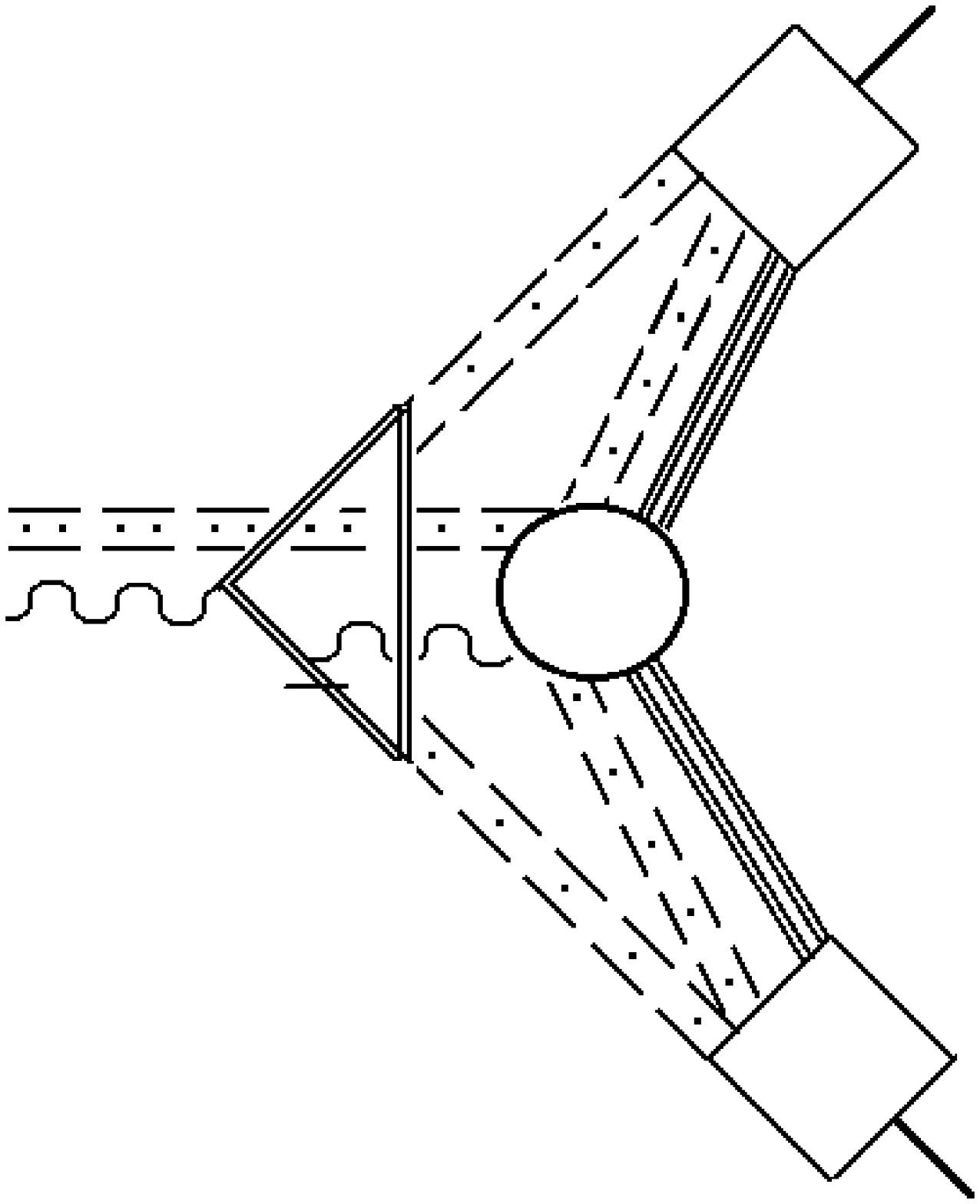}
\epsfxsize=3.8in
\epsfbox{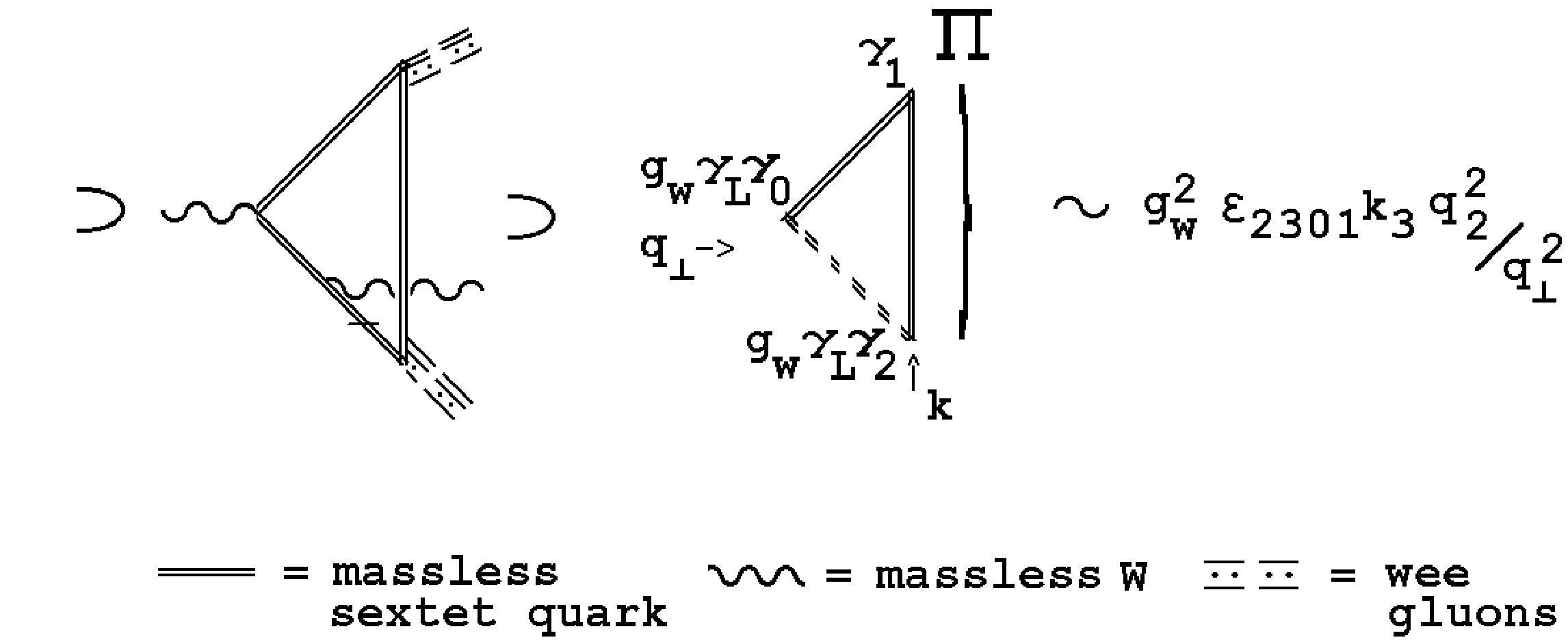}}
\caption{An Interaction Contributing to Vector Boson Mass Generation.}
\end{figure}
The sextet quark loop contribution is equivalent to the 
mixing\footnote{The anomaly chirality transition is necessary for this mixing.}
of a sextet pion ($\Pi$) 
with a $W$ and gives $M_W^2 ~\sim~ g_W^2 ~\int dk ~k ~~\equiv 
~ g_W^2 ~F_{\Pi}^2~~~\sim ~$
where $k$ is a wee gluon momentum.
The wee gluon origin implies 
this mass appears only in the S-Matrix. It also occurs only 
for vectors with a left-handed coupling. No photon (or gluon) mass 
is generated !

Both sextet and triplet pions  
contribute to the vector boson mass generation but, 
sextet pions dominate because of larger
color factors. Assuming the Casimir Scaling rule holds
($~ C_6~\alpha_s (F_{\Pi}^2)~\sim ~C_3 ~\alpha_s(F_{\pi}^2)$ with
$C_6/C_3 ~\approx~ 3~$), if $\alpha_s$ evolves sufficiently slowly ( e.g. $
\alpha_s (F_{\pi}^2) \sim 0.4~$) then
$F_{\Pi}$ can consistently be the electroweak scale! The large wee gluon 
coupling to sextet quarks also implies that
the wee gluon component of the pomeron
couples very strongly ($~\sim F_{\Pi}~$) to sextet quarks. 
This strong coupling is a central element of our understanding of high-energy sextet 
cross-sections. 

\section{New Strong Interaction Physics}

In general, the larger color factors involved imply that the sextet sector will 
be a higher mass, stronger coupling, part of the strong interaction.
As a result sextet states and cross-sections will dominate at high
enough energy. At first sight, the proposition that the sought-after 
new physics ``Beyond the Standard Model'' is strong interaction 
physics seems very radical. However, it has a number of theoretical 
attractions and potentially answers a number of outstanding 
experimental problems, as the following list\cite{arw04} illustrates.
{\openup-0.5\jot
\parindent=0.1in  
\begin{enumerate}
\item{Electroweak symmetry breaking involves no new interaction besides 
the $SU(3)\otimes SU(2)\otimes U(1)$ gauge 
interactions of the Standard Model.}
\item{The electroweak scale is a new QCD (sextet chiral) scale, 
which Casimir scaling shows to be the right order of magnitude.}
\item{The spectrum is more limited than just color confinement and 
chiral symmetry breaking would imply, in better accord with experiment.  
Glueballs and quark resonances are excluded as asymptotic 
states. There is no BFKL pomeron and no odderon. }
\item{New large cross-section physics above the electroweak scale gives 
a natural explanation for many cosmic ray phenomena. Big 
increases in average transverse momenta, and in the production of 
sextet neutrals, could give the apparent ``knee'' in the spectrum.}
\item{The high-energy production of stable, neutral, sextet neutrons, in the 
early universe, provides a 
natural explanation for the dominance of dark matter - formed as 
nuclei, clumps, etc. from sextet neutrons. }
\item{Being neutral and massive, 
sextet neutrons will avoid the GZK cut-off and so could 
be the mysterious, ultra-high energy, cosmic rays.}
\item{Top quark production could be due to the $\eta_6$
and sextet quark effects could explain an excess, large $E_T$, jet cross-section 
at the Tevatron.}
\end{enumerate}
}
\parindent=0.5in

\section{LHC Physics} 

If the cosmic ray spectrum knee is associated with an effective 
threshold for sextet quark physics (the energy is right,
given that pomeron production is needed), large 
cross-section effects must appear very rapidly with energy 
to make the knee visible. Such effects should be 
apparent at the LHC, with dramatic new physics involved\cite{arw04}. 
Considering anomaly pole amplitudes, we can show that the strong coupling 
of the pomeron to sextet states implies that hard double pomeron production 
of electroweak vector bosons will give jet cross-sections
comparable with normal QCD jet (non-diffractive) cross-sections. 
In particular, the boson pair cross-section is estimated to be, roughly, twelve 
orders of magnitude larger than in the Standard model. 
Combining this estimate with pomeron regge theory
gives a small transverse momentum cross-section that is 
correspondingly large. During the initial ``soft physics'' 
running period, it should be straightforward\cite{mga} to look for the vector 
boson pairs in the CMS central detector that should be produced
in combination with scattered protons in the TOTEM Roman Pots.

If pomeron exchange amplitudes are large, then cut-pomeron amplitudes should also be 
large. As a result, we expect very large 
inclusive cross-sections for sextet states (multiple vector bosons, 
in particular) across most of the rapidity axis. This implies 
that jet cross-sections, at very large transverse 
momentum, will be orders of magnitude larger than expected. 
The production cross-section for sextet nucleon pairs should also be 
hadronic in size, although stable sextet neutrons (dark matter!) may be 
difficult to detect. 
If the sextet nucleon double pomeron cross-section is 
extraordinarily large, it might be detectable in the low 
luminosity run. If not, it might be seen by the high luminosity 
detectors\cite{mga} that will look for double pomeron production of the 
Standard Model Higgs particle. 

\section { GUT$_S$ }

Well above the electroweak 
scale, the  $QCD_S$ infra-red fixed point requires 
$\alpha_s ~\centerunder{$<$}{$\sim$} 
~ \frac{1}{34}\sim \alpha_{ew} $,
implying that the sextet sector can, naturally, produce the decrease in $\alpha_s$ 
needed for unification. (Supersymmetry is not required !) A priori, 
unification could also determine how the $SU(2)\otimes U(1)$ sextet sector 
anomaly is canceled, as well as providing an origin for masses.

Many years ago (with Kyungsik Kang) we found\cite{kw} a remarkable, but puzzling, result. 
We looked at  asymptotically free, anomaly-free. left-handed 
unified theories that contain the sextet sector,
We discovered that a unique theory is selected, i.e.
$SU(5)$ gauge theory with the fermion representation $5+15+40+45^*$ 
($\equiv$ GUT$_S$). 
(In any higher unitary group the sextet sector requires
a representation that is too large for asymptotic freedom.) Under 
$SU(3)\otimes SU(2)\otimes U(1)$ 
\newline $~~~~$ 

\centerline{${\scriptstyle \bf 5~=~(1,3,-\frac{1}{3})~+~(2,1,\frac{1}{2})}~,~~~~ $
${\scriptstyle \bf 15~=~(1,3,1)~+~(3,2,\frac{1}{6})~+~{\bf \{(6,1,-\frac{2}{3})\}}}~,
~~~~~~~~~~~~~$}
\centerline{ ${\scriptstyle \bf 40~=~(1,2,-\frac{3}{2})~+~(3,2,\frac{1}{6})~+
~(3^*,1,-\frac{2}{3})~+~(3^*,3,-\frac{2}{3})~+~
{\bf \{(6^*,2,\frac{1}{6})\}}~+~(8,1,1)}~,
~~~~~~~~~$}
\centerline{${\scriptstyle \bf 45^*=~(1,2,-\frac{1}{2})~+~(3^*,1,\frac{1}{3})
~+~(3^*,3,\frac{1}{3})~+~(3,1,-\frac{4}{3})~+~(3,2,\frac{7}{6})~+~
{\bf \{(6,1,\frac{1}{3})\}} ~+~(8,2,-\frac{1}{2})}$}

\noindent The triplet quark and lepton sectors, although not asked for, 
are remarkably close
to the Standard Model. There are three ``generations'' of 
quarks/anti-quarks, with quark charges $\frac{2}{3}$ and $-\frac{1}{3}$, 
and  three ``generations'' of $SU(2)$ doublet 
($SU(3)$ singlet) leptons. The puzzle is  
that the $SU(2)\otimes U(1)$ quantum 
numbers are almost, but not quite, right and there are also 
(apparently unwanted) color octet quarks with lepton-like
electroweak quantum numbers. At the time\cite{kw}, we considered various ``anomalous
fermion phenomena'', but found no convincing dynamical route
to the Standard Model.
 
\section{ A Massless Theory of Matter ? }

In fact, GUT$_S$ has similar properties to massless $QCD_S$. Asymptotic freedom is 
saturated and an infra-red fixed-point keeps the $SU(5)$ coupling very small. 
Also, the $SU(5)$ symmetry can be broken to $SU(4)$ 
with an asymptotically free scalar field and so the high-energy S-Matrix can 
be constructed via reggeon diagram anomaly interactions introduced by the symmetry
breaking. Because infra-red divergences 
will confine $SU(5)$ color in the S-Matrix, {\it all elementary
fermions will be massless and confined}. The  
massless Dirac sea will control the dynamics, but with 
a crucial difference from QCD$_S$. In GUT$_S$, left-handed fermion
interactions will exponentiate the anomalous color parity divergences 
that produce the states and amplitudes of QCD$_S$. 
Therefore, these divergences {\it will only be produced by  
the  $SU(3)\otimes U(1)$ vector part of the theory.} The left-handed vector bosons,
with no SU(3) color, will aquire a mass, as in Section 8. Thus, just the 
interactions of the Standard Model will be selected within the GUT$_S$ S-Matrix.

As yet, very little is certain, but our current ideas\cite{arw05} about the S-Matrix 
construction, and
other properties of GUT$_S$, can be listed as follows.
(We use {\small $ ~SU(5)~\to~SU(3)_C\otimes SU(2)_L\otimes U(1)
~\to~SU(2)_C\otimes SU(2)_L\otimes U(1)$}.)
{\openup-1\jot
\begin{itemize}
\item{\it We begin 
with a $k_{\perp}$ cut-off and SU(5) broken to SU(2)$_C$.} 
\item{\it The states are SU(2)$_C$ singlet Goldstones ($~\pi$'s~) which 
are $q\bar{q}$ pairs in an SU(2)$_C$ condensate, as in CSQCD$_S$.}
\item{\it The ~$\pi$'s~ 
contain $q$'s that are { \bf 3's, 6's,} and {\bf 8's} under SU(3)$_C$.
\newline ({\bf 8's} are real wrt SU(3)$_C$,  
but give complex doublets wrt SU(2)$_C$.)}
\item{\it Anomaly interactions, involving the
SU(2)$_C$ condensate, generate $W^{\pm}$ and $Z^0$ masses via mixing (predominantly)
with sextet $\pi$'s.}
\item{\it Similar interactions should generate $\pi$ masses. The pattern,
and how many parameters are involved, remains to be studied.} 
\item{{\it Restoring SU(2)$_L$ symmetry gives
SU(2)$_C$} x {\it SU(2)$_L$ invariant states.}}
\item{\it ``Octet pions'' form bound state leptons with 
elementary ``leptons''.}
\item{\it After SU(2)$_L$ confinement,
the right-handed sextet flavor dependence gives a (broken) SU(2) 
S-Matrix symmetry. }
\item{\it With SU(3)$_C$ color, 
the Pomeron is Critical and the photon massless.}
\item{\it SU(3) octet quarks have no chiral anomaly and hence no pomeron 
coupling - hence {\it no strong interaction!} They appear in bound-state leptons 
as pairs of positive and negative energy massive states.} 
\item{\it Because the coupling is so small, the underlying SU(5) theory will be
close to conformal and the smallest lepton masses will be very small.}
\item{\it There are no spectrum symmetries that would prevent the
S-matrix from having the full structure of the Standard Model.}
\item{\it The (experimentally attractive) SU(5) value of the 
Weinberg angle should be valid for GUT$_S$, even though
there is no proton decay !}
\item{\it The {\bf 144} representation of SO(10) is
anomaly free, asymptotically free, and contains GUT$_S$. There is, however, no
``saturation'' and so we can not construct a unitary, anomaly based, 
high-energy S-Matrix.} 
\end{itemize}
}

\section{Comments}

The success of the Standard Model is a major 
scientific achievement. 
Nevertheless, it remains an enigma in the sense that there are a large 
number of parameters and there is no 
understanding of why it has been chosen by nature. 
The candidate theories/ideas that have so far been proposed
for extending the model (including supersymmetry and superstrings) have 
failed to provide any understanding of the model's uniqueness and, as a result, 
have provided a wide range of possible extensions.

In the search to understand the origin of the 
Standard Model, the importance of asking that 
physical states give a unitary S-Matrix may been
underestimated. When S-Matrix theory reigned supreme\cite{arw00}, 
before the dramatic rebirth of field theory that has led to the 
Standard Model, awareness of the difficulty involved led to the 
proposal that there is a unique S-Matrix that could, in principle,
be bootstrapped from unitarity. 

In an asymptotically
free field theory, the wild divergence (non-summability) of  
perturbative interactions occurs
in the infra-red region. Presumably, it is very difficult to keep 
this divergence out of the S-Matrix. Perhaps, there has to be
a drastic reduction 
of the field theory degrees of freedom, in the states and in the interactions,
that is only achievable (in very special circumstances)
via infra-red anomalies of the massless Dirac sea. Could it be that 
the unitary S-Matrix is indeed unique but that to construct it, knowledge of 
the (also unique) underlying {\it massless field theory} is essential?

\end{document}